\documentclass[letter]{aa}  
\usepackage{graphicx}
\usepackage{txfonts}
\begin{document}

\title{The possible evolution of pitch angles of spiral galaxies}

\author{V.P. Reshetnikov\inst{1,2}, A.A. Marchuk\inst{1,2}, I.V. Chugunov\inst{1,2},
           P.A. Usachev\inst{1,2}
          \and
          A.V. Mosenkov\inst{3}
          }

\institute{St. Petersburg State University, Universitetskii pr. 28, St. Petersburg, 198504 Russia
         \and
Pulkovo Astronomical Observatory, Russian Academy of Sciences, St. Petersburg, 196140 Russia
         \and
Department of Physics and Astronomy, N283 ESC, Brigham Young University, Provo, UT 84602, USA
             }

\date{Received October , 2023; accepted , 2023}

\abstract
{}
   {The origin and maintenance of spiral structure in galaxies, the correlation 
   between different types of spiral structure and several proposed mechanisms for their 
   generation, and the evolution of spiral arms of 
   galaxies with time are questions that are still  controversial. In this note we study the 
   spiral structure in a sample of distant galaxies in order to infer the evolution of 
   spiral arm characteristics with redshift.}
   {We considered a sample of 171 face-on spiral galaxies in the Hubble
   Space Telescope COSMOS (The Cosmic Evolution Survey) field. The galaxies are distributed 
   up to $z \approx 1$ with a mean value of 0.44. For all galaxies, we determined the pitch 
   angles of the spiral arms and analysed their dependence on redshift; 
   a total of 359 arms were measured.}
   {Analyses of our measurements combined with the literature data suggest a possible 
   evolution of the pitch angles of spiral galaxies:  by the modern epoch the spiral 
   pattern, on average, becomes more tightly wound. This may be a consequence of the 
   general evolution of the structure of galaxies as galaxies become more massive 
   over time and their bulges grow. In addition, the distribution of the cotangent of
   pitch angle of galaxies indicates the possibility that the dominant mechanism
   of spiral pattern generation changes over time. }
   {}

\keywords{galaxies: structure -- galaxies: spiral -- galaxies: evolution
               }

\maketitle
\nolinenumbers

%

\section{Introduction}

Spiral structure is the most prominent and longest recognized (e.g. \citealt{rosse})
distinctive feature of the galaxies around us. Spiral arms shape what a galaxy looks 
like, so they are traditionally used to morphologically classify galaxies (\citealt{hubble}).
A spiral pattern is present in most nearby massive galaxies (e.g.
\citealt{nair}; \citealt{willett}), although the mechanisms through
which these spiral structures originate and evolve are still debated (e.g. \citealt{doba}).
It appears that the nature and dominant formation mechanisms of spiral arms may vary from one
galaxy to another.

A substantial amount of observational data on the structure of spiral arms in nearby 
galaxies has now been accumulated (see discussion and references in \citealt{ARAA2022}).
On the other hand, systematic studies of the spiral pattern in distant galaxies 
are still very rare (e.g. \citealt{savch2011}; \citealt{davis}; \citealt{martinez2023}). 
\citet{ee2014} showed that all types of spiral patterns known in the nearby Universe (grand-design, 
flocculent, and multiple arms) are present at $z \geq 1$. Galaxies with pronounced 
spiral arms are also occasionally detected at high redshifts, for instance 
$z=2.54$ (\citealt{yuan2017}), $z=3.06$ (\citealt{wu2023}), and $z=4.41$ (\citealt{tsukui}).
Recent studies suggest that high $z$ may be dominated by disc galaxies (e.g. \citealt{ferreira}). 
This suggests that spiral pattern in distant galaxies may be quite common. 

In our previous work (\citealt{resh}, hereafter Paper I), we present the results of 
determining the
spiral arm pitch angles for 102 galaxies with a two-armed pattern in the 
Hubble Space Telescope (HST) Cosmic Evolution Survey (COSMOS) field. 
We found a weak but 
distinct tendency for the pitch angle to decrease as we approach the present-day epoch. 
Furthermore, our analysis using the Pringle-Dobbs test (\citealt{pringle}) showed that the
predominant mechanisms of spiral arm formation can change over time (Paper I).
In this note we almost double the number of distant galaxies studied and present the  
results of exploring the possible evolution of spiral pattern of galaxies with time.

Throughout this article we adopt a standard flat $\Lambda$CDM
cosmology with $\Omega_m$=0.3, $\Omega_{\Lambda}$=0.7, and  $H_0$=70 km\,s$^{-1}$\,Mpc$^{-1}$.


\section{Sample and data reduction}

Galaxies for this study were selected from the HST/Advanced Camera for Surveys COSMOS field.
This survey covers almost 2\,deg$^2$ and provides deep and high-resolution 
F814W imaging data (\citealt{koek}). As in Paper I, to select objects with spiral
arms we used the same sample comprising 26113 bright (F814W < $22\fm5$) galaxies in the COSMOS
field presented by \citet{mand}. In the first step the apparent flattenings ($b/a$) 
were determined for each galaxy using the {\sc sextractor} package (\citealt{sextr}) 
and galaxies with $b/a > 0.7$ were selected.   Based on careful visual 
inspection, a final sample was then selected consisting   of 171 galaxies with clearly 
discernible and measurable spiral branches; the pixel size of the images is $0\farcs03$. 
The mean axial ratio of the galaxies in the final sample is 
$\langle b/a  \rangle = 0.87 \pm 0.06$. In the following we assume that all galaxies
are seen exactly face-on and do not use any correction for the possible inclination effect.

\begin{figure}
\centering
\includegraphics[width=11.5cm, angle=-90, clip=]{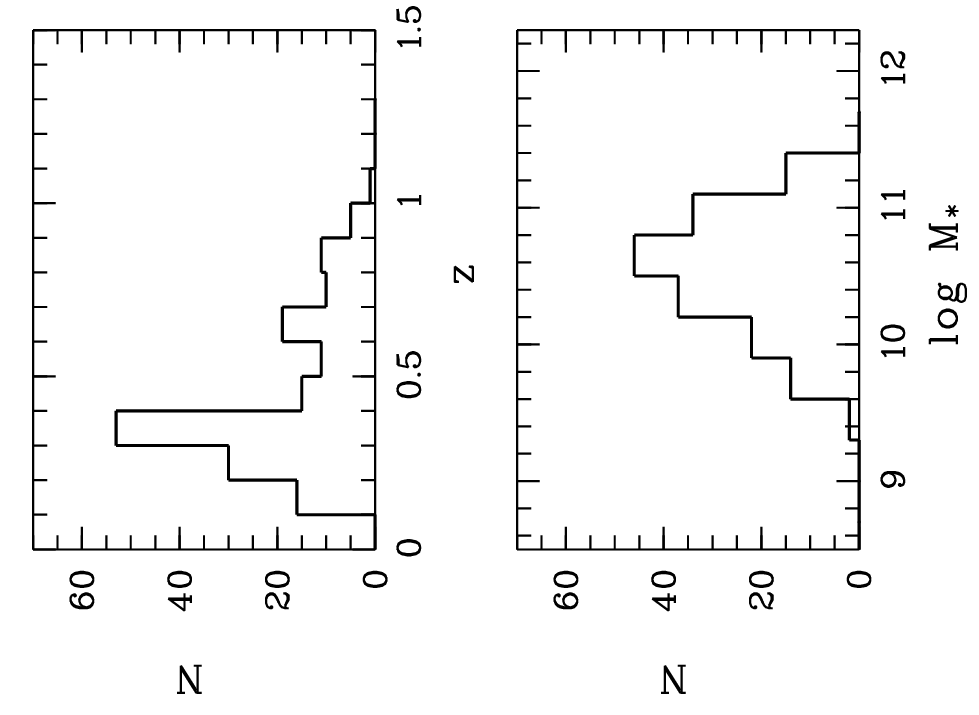}
\caption{Distributions of redshift (top) and stellar mass in solar units (bottom) for 
the sample galaxies.}
\label{fig:hist}
\end{figure}

The sample objects were identified with the COSMOS2020 catalogue (\citealt{weaver}).
For each galaxy, we adopted the photometric redshift and stellar mass found
with the LePhare code (\citealt{ilbert}). \autoref{fig:hist} illustrates the main 
characteristics of the sample: galaxies are distributed up to $z\approx1$ (mean redshift
$\langle z \rangle = 0.44 \pm 0.23$); galaxy stellar masses are contained within 
$9.5\,\leq\,$\,log\,M$_*$/M$_{\odot}\,\leq$\,11.5 with a mean value of 10.53$\pm$0.43.

\begin{figure*}
\centering
\includegraphics[width=7cm, angle=0, clip=]{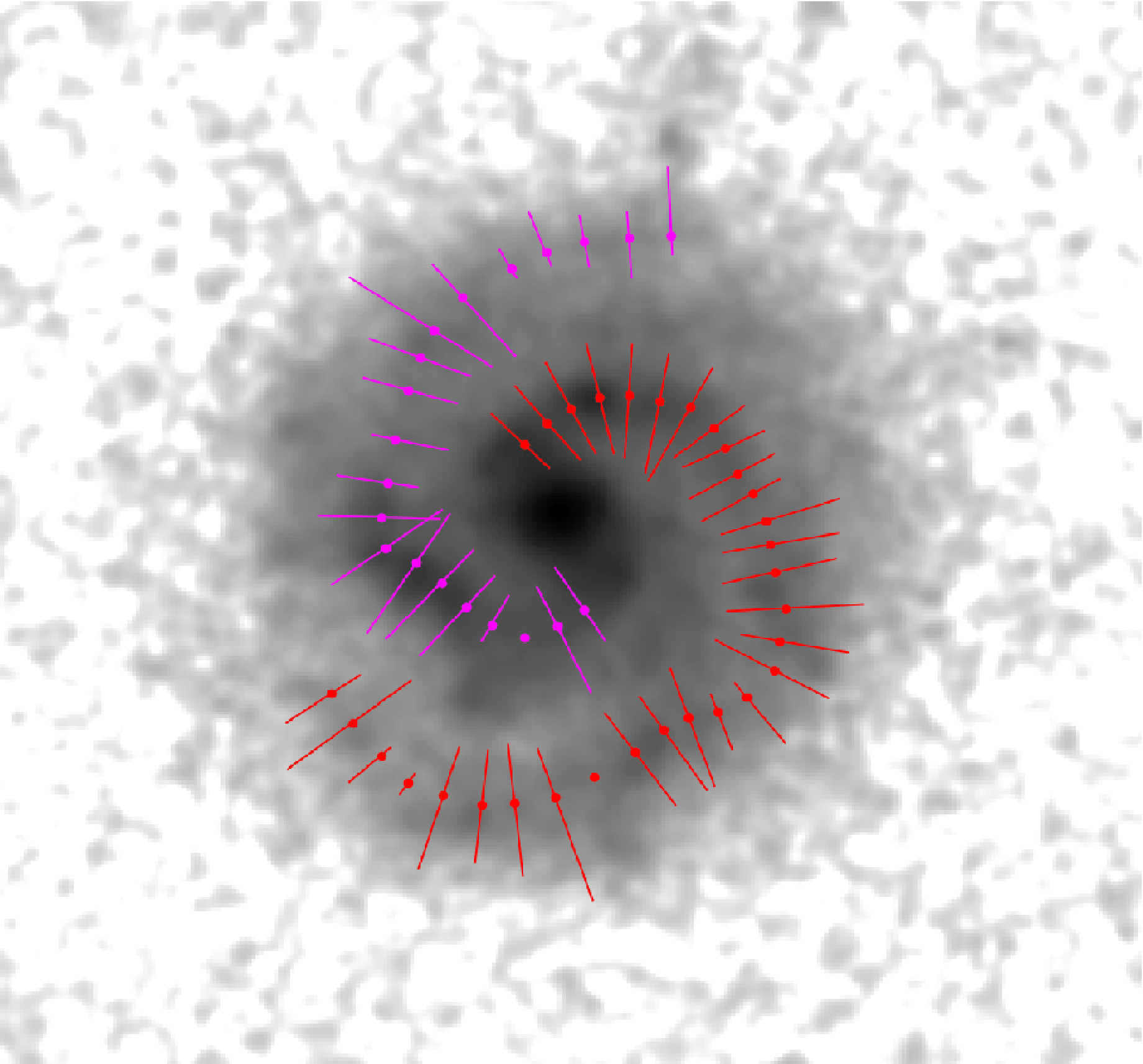}~~
\includegraphics[width=7cm, angle=0, clip=]{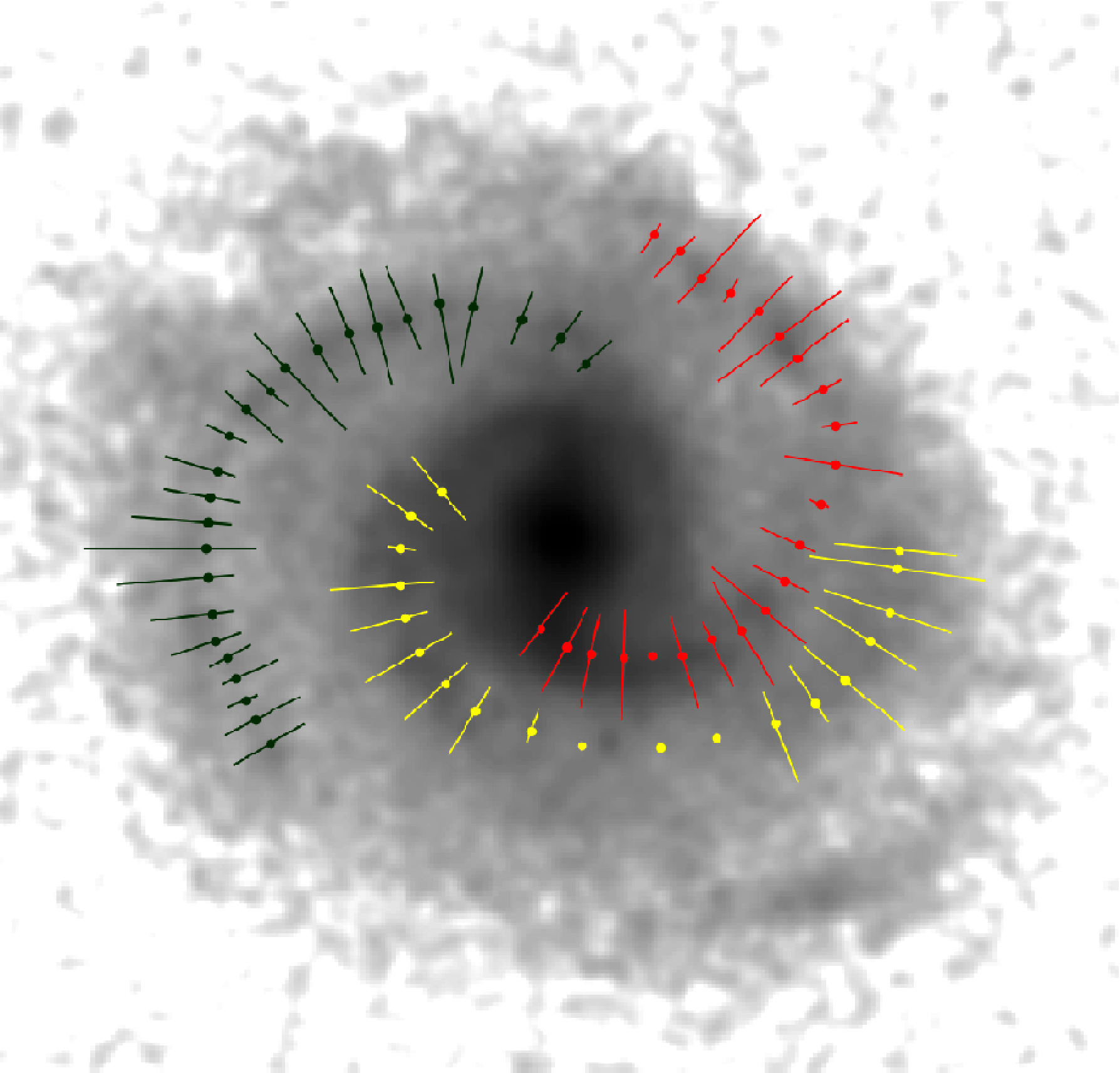}
\caption{Examples of galaxies from the sample with measured spirals 
(left: ra=150.540631 dec=2.169663, $z = 0.31$; 
right: ra=150.500902 dec=2.611881, $z = 0.38$;  see \citealt{weaver}). 
The dots give the positions of the brightness peaks; the segments correspond to 
the spiral half-width.}
\label{fig:images}
\end{figure*}

The procedure for measuring the pitch angles of spiral arms was the same as in Paper I
(see also \citealt{savch2020}). The number of measured arms in one galaxy varied
from 1 to 5: 29 galaxies with 1 measured arm, 104 with 2 arms, 31 with 3 arms,
6 with 4 arms, and 1  with 5 arms (see examples in \autoref{fig:images}). 
In total, we determined the pitch angles of 359 arms in 171 galaxies. 

\begin{figure}
\centering
\includegraphics[width=6.5cm, angle=-90, clip=]{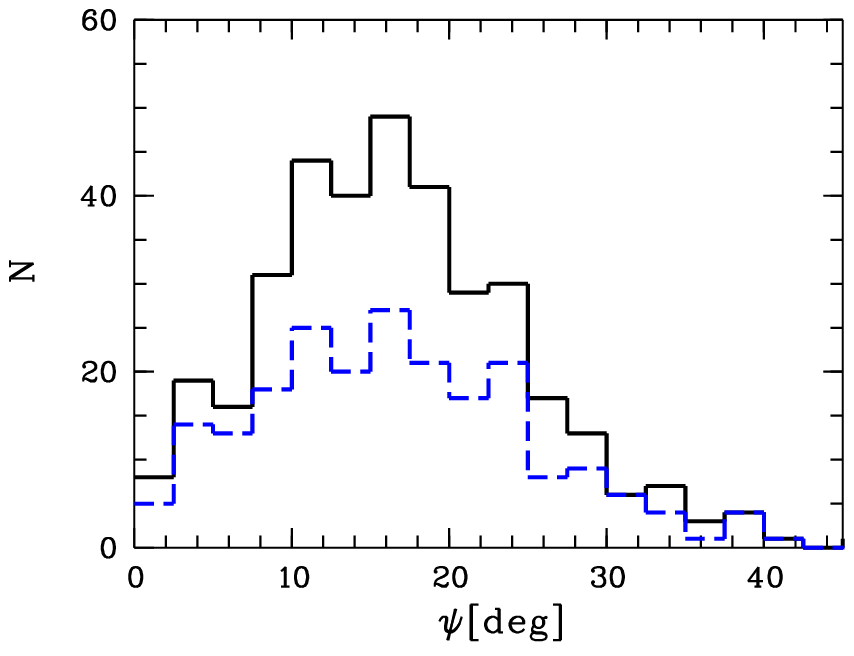}
\caption{Distribution of spiral arm pitch angle $\psi$ for the sample galaxies
(solid line). The dashed line shows the distribution for massive galaxies with
log\,M$_*$/M$_{\odot} \geq 10.5$. }
\label{fig:pitch}
\end{figure}

The mean pitch angle of all measured arms is
$\langle \psi \rangle = 16.^{\circ}65 \pm 8.^{\circ}17$; the same value for massive
galaxies with log\,M$_*$/M$_{\odot} \geq 10.5$ is 
$\langle \psi \rangle = 16.^{\circ}64 \pm 8.^{\circ}62$ (214 arms).
The mean error of the pitch angle measurement is $\langle \sigma_{\psi} \rangle = 1.^{\circ}6$ 
(see also fig. 4 in Paper I). Within the observed scatter of the pitch angles
there is no statistically significant difference in the 
mean $\psi$ values for galaxies with different numbers of spiral arms.
As seen in \autoref{fig:pitch}, the angles of galaxies at $z \leq 1$, similar to nearby
objects (e.g. \citealt{yu2020}), are distributed mainly within $\psi \leq 40^{\circ}$.

\section{Pitch angle evolution}

\subsection{$\psi$--$z$ correlation}

\begin{figure}
\centering
\includegraphics[width=6.0cm, angle=-90, clip=]{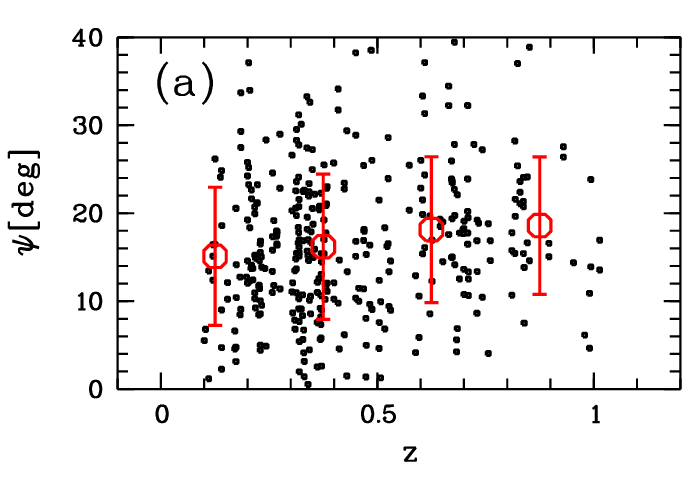}
\includegraphics[width=6.0cm, angle=-90, clip=]{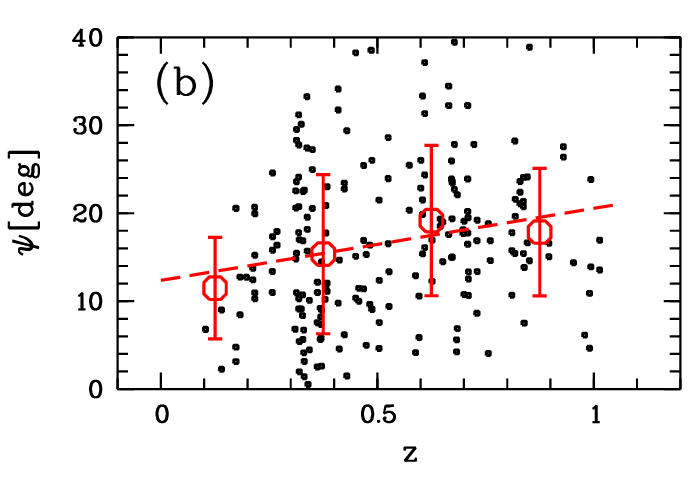}
\caption{Redshift dependence of pitch angle:
(a) entire sample, (b) massive galaxies with log\,M$_*$/M$_{\odot} \geq 10.5$.
The circles with bars give the mean values in redshift bins 0--0.25, 0.25--0.50,
0.50--0.75, 0.75--1. The dashed line is the linear regression for massive spiral
galaxies: 
$\psi(^{\circ}) = (8.^{\circ}2 \pm 2.^{\circ}6)\cdot z + (12.^{\circ}4 \pm 1.^{\circ}5)$.}
\label{fig:evol}
\end{figure}

In \autoref{fig:evol}a,b, we consider the dependence of the pitch angle of 
individual spiral arms on redshift. The scatter of the data is large, but 
a slight tendency, a weak trend, is apparent: as $z$ increases, 
$\psi$ increases as well. The correlations in \autoref{fig:evol}a,b are weak
(the coefficients of linear regression are about 0.2), but given 
the relatively large number of objects, they are significant at $> 95\%$.

We point out that the pitch angles of galaxies are measured with errors. 
They show natural variations with radius (e.g. \citealt{savch2013}) and, in addition, 
different arms in the same galaxy 
may have different average pitch angles (e.g. \citealt{savch2020}; Paper I). 
Therefore, the question arises of how strong the correlation with the pitch angle can be if it
is determined with a large ($\sim50\%$) error.

To illustrate the effect of errors on the observed correlation, we performed a simple 
numerical experiment. We considered a pure linear dependence (the coefficient of linear
correlation is 1.0) of two independent variables. The number of points was taken 
as equal to the number of spiral arms measured in our sample (N=359). Then, a random error 
distributed according to the normal law was introduced into one of the variables, with 
the dispersion equal to 50\% of the mean value of the variable, and the linear regression
coefficient of such an artificially noisy dependence was measured again. After performing this 
operation many times, we obtained that after the introduction of errors, the correlation 
coefficient decreased to about 0.75. For the linear correlation coefficient to be equal 
to 0.2, as in our sample, the correlation coefficient of the original, unnoised dependence 
should be 0.4--0.5. We note that if the error is not normally distributed, the reduction 
in the observed correlation may be even stronger. Thus, we can conclude that the current 
data on the pitch angles of distant galaxies do not contradict the existence of a moderate 
dependence of $\psi$ on redshift.

\subsection{The Pringle--Dobbs test}

For transient and recurrent spiral arms, and for tidal spiral arms, \citet{pringle} proposed 
that, under some simple assumptions, the pitch angle $\psi$ decrease systematically with 
time $t$ as $cot\,\psi \propto t$.
This assumption is consistent with numerical simulations of both the transient and
the tidal arms (e.g. \citealt{grand}; \citealt{pettitt}).
This means that
if we observe a random sample of galaxies at any fixed time, we expect to see a uniform
distribution of $cot\,\psi$ values. Application of this test to nearby galaxies
showed that the distribution of $cot\,\psi$ values is indeed close to uniform
(\citealt{pringle}; \citealt{lingard}). 

\begin{figure}
\centering
\includegraphics[width=16cm, angle=-90, clip=]{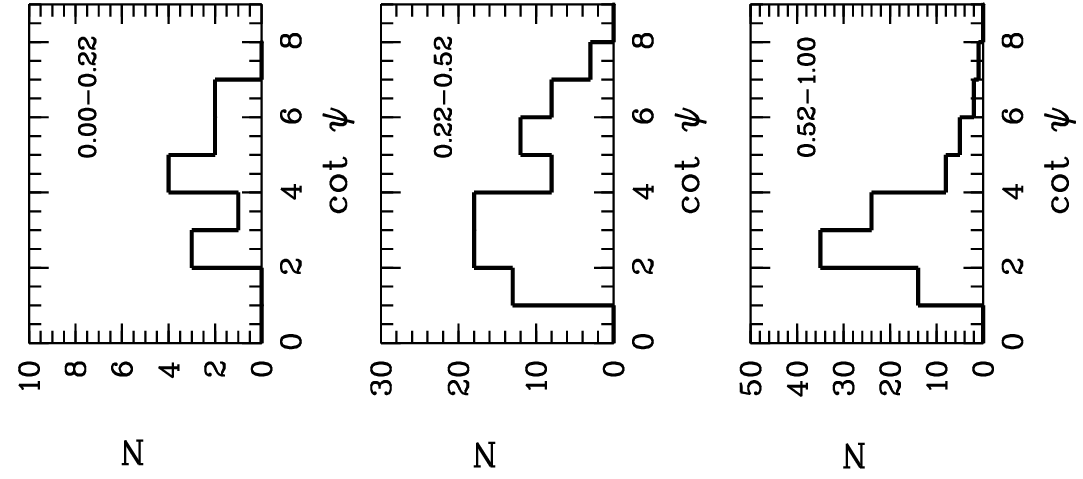}
\caption{Distributions of $cot\,\psi$ for the spiral arms of massive galaxies 
(log\,M$_*$/M$_{\odot} \geq 10.5$) in different redshift bins.}
\label{fig:cot}
\end{figure}

\autoref{fig:cot} illustrates the Pringle--Dobs test for
massive spirals in three redshift bins. The sizes of the bins were chosen to correspond
to approximately the same step in time $\Delta t = 2.6$ Gyr. In the first bin ($z \leq 0.22$)
there are few galaxies,  and they show an almost flat distribution (see also fig.\,7 in Paper I).
The statistics are small, but not inconsistent with the results of \citet{pringle} and
\citet{lingard} for nearby spiral galaxies. In the second bin ($z = 0.22 - 0.52$), 
the galaxies are much more numerous and show a broad and relatively flat distribution. 
In the more distant galaxies ($z = 0.52 - 1.00$) the appearance of the distribution changes: 
the right wing is suppressed; $cot\,\psi$ shows a pronounced single peak at $cot\,\psi
\approx 2.5$ ($\psi \approx 22^{\circ}$). This change in the type of the distribution 
may reflect a change in the dominant mechanism of the generation and maintenance of 
spiral pattern in different epochs.

\subsection{Pitch angles of most distant galaxies}

\begin{figure}
\centering
\includegraphics[width=6.0cm, angle=-90, clip=]{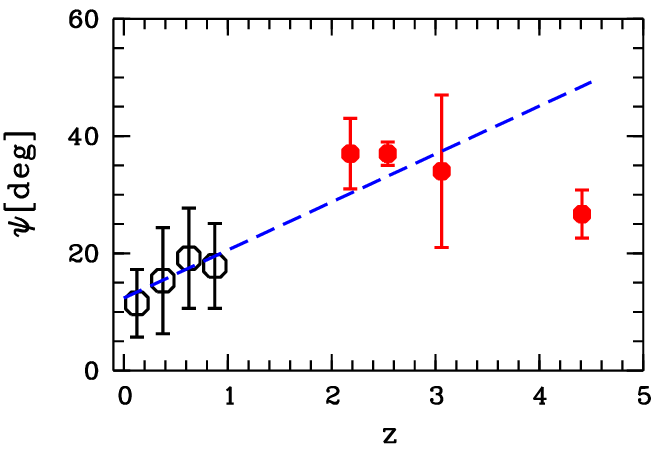}
\caption{Dependence of pitch angle of spiral galaxies on redshift. The open cirles depict
mean values for the COSMOS spirals (see \autoref{fig:evol}b); the solid circles are the 
individual measurements for several distant galaxies (see  text). The dashed line
shows an extrapolation of the linear regression from \autoref{fig:evol}b to high redshifts.}
\label{fig:evolz}
\end{figure}

We now consider how the trend shown in \autoref{fig:evol} agrees with 
observations of high-redshift galaxies. Unfortunately, the data on the pitch angles
of the spiral arms of distant galaxies are very sparse to date.
\autoref{fig:evolz} shows the characteristics of spiral pattern of four
galaxies with $z > 2$. These objects are as follows: 

Q2343-BX442, $z = 2.18$, $\psi = 37^{\circ} \pm 6^{\circ}$ (\citealt{law2012});

A1689B11, $z = 2.54$, $\psi = 37^{\circ} \pm 2^{\circ}$ (\citealt{yuan2017});

A2744-DSG-z3, $z = 3.06$, $\psi = 34^{\circ} \pm 13^{\circ}$ (\citealt{wu2023});

BRI~1335-0417, $z = 4.41$, $\psi = 26.^{\circ}7^{+4.^{\circ}1}_{-1.^{\circ}6}$ (\citealt{tsukui}).

As can be seen in \autoref{fig:evolz}, the pitch angles of distant galaxies exceed the 
values for nearby objects and, in general, they agree with the observational trend noted in 
\autoref{fig:evol}. The statistics of distant objects are, of course, exceptionally 
poor, so \autoref{fig:evolz}   merely characterises the current state of affairs with 
measurements of spirals at $z > 2$.

\section{Conclusions}

In this note we have presented the results of determining the pitch angles of the spiral arms 
in 171 spiral galaxies up to $z \approx 1$ in the HST COSMOS field. Our measurements
were performed in the F814W filter, which approximately corresponds to the rest-frame 
$V$ filter at the mean redshift of the sample. 

Based on our data, and using literature data on the structure of the most distant galaxies 
with spiral pattern measurements, we conclude that there may be an evolution in the shape of 
spiral arms with time. As we approach the modern epoch, spiral arms, on average, become more tightly
wound. The amplitude of the effect is small and corresponds to  $\sim 1^{\circ}$/Gyr
at $z \leq 1$.

Tidal spirals and recurrent transient spiral instabilities, driven by self-gravity,
evolve with time such that the pitch angle of the spiral pattern decreases (e.g. 
\citealt{grand}; \citealt{semczuk}; \citealt{pettitt}). However,  this twisting occurs on
a much smaller timescale and with greater amplitude. In our letter we discuss
the longer overall evolution of the pitch angle.

A possible reason for the observed long-term evolution of the spiral pitch angle may 
be the evolution of the global structure of galaxies. Over time, galaxies become more 
massive, their concentration towards the centre rises, and bulges become progressively
brighter. More than half of the  current stellar mass of the bulges is gained at
$z \lesssim 1$ (\citealt{sachdeva}). On the other hand, a number of studies have
shown that the pitch angle correlates with the mass of the galaxy and the characteristics
of its bulge (e.g. \citealt{savch2013}; \citealt{yu2019,yu2020}).
Consequently, the change in the structure of evolving galaxies should be accompanied 
by a gradual change in the shape of the forming spiral arms.
In addition, the effects of  observational selection, cosmological surface brightness 
dimming, and $k$-correction  can all influence the observed correlation.

For objects at different redshifts, we considered a distribution  of $cot\,\psi$.
Within the current relatively sparse statistics we can tentatively conclude that 
at $z < 0.5$ the shape of the $cot\,\psi$ distribution changes (see also Paper I). 
This may indicate that the dominant mechanism for generating and maintaining the spiral 
pattern changed during that epoch (about 5 Gyr ago) and the surrounding part of the
Universe is dominated by transient and recurrent spiral arms driven by
self-gravity (\citealt{pringle}; \citealt{lingard}). 

The results of this work are based on the Hubble Space Telescope imaging data. 
The James Webb Space Telescope (JWST) significantly outperforms the HST in its ability to highlight spiral structure in 
distant and faint galaxies (see e.g. Fig.\,14 in \citealt{ferreira}).
Therefore, future observations with the JWST will hopefully clarify the issues
discussed in this note.

\begin{acknowledgements}
This work was supported by the Russian Science Foundation (project no. 22-22-00483).
\end{acknowledgements}

\end{document}